\documentclass[preprint,nofootinbib,aps,superscriptaddress,eqsecnum]{revtex4-1}
 \pdfoutput=1
\usepackage{cancel}

\usepackage{float}
\usepackage{graphicx}
\usepackage{amsmath}
\usepackage{amsfonts}
\usepackage{amssymb}
\usepackage[bookmarks=true,bookmarksopen=true,
                    bookmarksnumbered=true,
                    colorlinks=true,linkcolor=black,
                    citecolor=red]{hyperref}
\usepackage{subfigure}
\usepackage{slashed}
 \usepackage{setspace} 
\usepackage{caption}
\usepackage{xcolor}
\allowdisplaybreaks
\def\bea{\begin{eqnarray}}
\def\eea{\end{eqnarray}}
\def\be{\begin{equation}}
\def\ee{\end{equation}}

\begin{document}
\title{$\texttt{SpaceMath v.2.0}$ with Machine Learning.\\
		A $\texttt{Mathematica}$ package for Beyond the Standard Model parameter space searches.	}

\author{M. A. Arroyo-Ure\~na}
\email{marco.arroyo@fcfm.buap.mx}
\affiliation{Facultad de ciencias f\'isico-matem\'aticas, \\
Benem\'erita Universidad Aut\'onoma de Puebla,\\
 C.P. 72570, Puebla, Pue., Mexico.}
 \affiliation{Centro Interdisciplinario de Investigaci\'on y Ense\~nanza de la Ciencia,\\
Benem\'erita Universidad Aut\'onoma de Puebla,\\
 C.P. 72570, Puebla, Pue., Mexico.}
 
 \author{T. A. Valencia-P\'erez}
 \email{tvalencia@fisica.unam.mx}
\affiliation{\small{Instituto de F\'isica\\
Universidad Nacional Aut\'onoma de M\'exico, C.P. 01000, CDMX, M\'exico.}}

 \begin{abstract}
	
	\texttt{SpaceMath v.2.0} with Machine Learning is an extension of the previous version  which we implement observables related with LHC Higgs boson data and their projections for the High Luminosity and High Energy Large Hadron Collider. In this version we implemented processes with Flavor-Changing Neutral Currents at tree and one-loop level, namely, i) Radiative decays $\ell_i\to\ell_j \gamma$, ii) $\ell_i\to \ell_j \ell_k \bar{\ell}_k$ decays ($\ell_i=\tau,\,\mu$, $\ell_{j,\,k}=\mu,\,e$, with $\ell_i    \neq\ell_j \neq\ell_k$) and iii) anomalous magnetic dipole moment of the muon $\delta a_\mu$. \texttt{SpaceMath v.2.0}  is able to find allowed regions for free parameters of models with both real and complex singlets and real and complex doublets using the processes previously mentioned within a friendly interface and an intuitive environment in which the user enters the couplings symbolically, sets parameters and execute \texttt{Mathematica} in the traditional way. As result, both tables as plots with values and areas agree with experimental data are generated. We present examples using \texttt{SpaceMath v.2.0} to analyze the free \textit{Two-Higgs Doublet Model of type III} parameter space, step by step, in order to start new users in a fast and efficient way. Finally, we have implemented in this version of \texttt{SpaceMath} algorithms of Machine Learning to generate specific Benchmark Points to be used directly in numerical evaluations of calculations of physical observables.
	\end{abstract}

\maketitle

	\markboth{F. Author \& S. Author}
	{Insert running title (not more than 8 words)}
	
	%
	%
	
%
%



	\tableofcontents 
	
	
%
	
	\section{Introduction}

	Our current knowledge of elementary particles and their interactions is based on solid theoretical foundations that are embodied in the Standard Model (SM). This theory provides a description of the weak, strong and electromagnetic interactions, satisfactorily explaining the experimental data, except for isolated cases. However, despite these achievements, there are phenomena that do not help us understand, for example: the problem of hierarchy, the origin of dark matter, the problem of flavor, etc. The fact that the SM cannot provide an answer to these phenomena suggests physics Beyond the SM (BSM). In the last decades, several extensions of the SM have been presented to try to solve them, this results in the emergence of free parameters that are not predicted by theory, though.
	
	The search for physics BSM is necessarily a multidisciplinary effort, since the evidence for new physics could appear in physical observables that have been proposed both theoretically and experimentally. One strategy is to produce new hypothetical particles in colliders (for example, LHC and future stages of it), searching in decays and/or in high precision measurements. In this context, the reports by different collaborations have given exclusions on specific regions of the parameter space that, however, have been valuable so far. On the other hand, there have been many supposed signatures of new physics, often only to be refuted by the lack of correlated signals in other experiments. Properly and fully weighing the sum of data relevant to a theory and making rigorous statistical statements about which models are allowed and which are not, has become a challenging task for both theory and experiment.
	
	Secondly, with the discovery of the Higgs boson \cite{HiggsDiscoveryATLAS, HiggsDiscoveryCMS} is established that the Higgs mechanism explains the electroweak symmetry breaking and it generate the mass of all particles of the SM, omitting the neutrino masses. However, it is well known that, despite its great success, the SM cannot help us to understand several issues, it encourages the study of SM extensions \cite{ArkaniHamed:2002qy, ArkaniHamed:2001nc, FRAMPTON1987157, GEORGI1985463, Harari:1979gi, Harari:1981uh, book:1299422, 10.1143/PTP.36.1266, PhysRevD.10.275, Mohapatra:1974hk, POLYAKOV1977429, Randall:1999ee, PhysRevD.20.2619, PhysRevD.19.1277, Arroyo-Urena:2019zah, Arroyo-Urena:2019lzv}, with the aim of solving some issue unexplained. The price to pay is the emergence of free parameters whose values are not predicted by the theory. From a phenomenological point of view, one frequently encounters these free parameters which should be constrained in some way, but at same time motivated and allowed by experimental measurements or by theoretical restrictions. With the $\texttt{SpaceMath}$ package, it is possible to do it. Free model parameter spaces can be constrained automatically within a friendly interface and an intuitive environment, where the user defines the couplings and executing the commands of $\texttt{SpaceMath}$ to generate both plots and tables showing the areas and numerical values according to experimental data. Similar packages to $\texttt{SpaceMath}$ can be consulted in the Refs. \cite{EasyScanHEP, GAMBIT, CheckMATE, Muhlleitner:2020wwk, Djouadi:2018xqq, DeBlas:2019ehy, Flacher:2008zq}. However, \texttt{SpaceMath} has the feature that it only requires the installation of \texttt{Wolfram Mathematica} (available in many universities and research institutes) and a very basic knowledge of \texttt{Wolfram} language. Unlike other programs that require prior knowledge of programming languages, the \texttt{SpaceMath} package has a fast learning curve and a practical approach which makes it an option for quick results.

	The organization of our work is as follows. In Sec. \ref{SeccionII} we present the theoretical framework necessary to have the basis of the programming of \texttt{SpaceMath}. Section \ref{SeccionIII} shows, in a concise way, the way to install \texttt{SpaceMath v.2.0} and show how it works, giving a detailed example. Sec. \ref{SeccionIV} is focused on the validation of \texttt{SpaceMath v.2.0} by reproducing several results shown in the literature. Finally, conclusion and perspectives are presented in Sec. \ref{SeccionV}.

\section{A theoretical overview}\label{SeccionII}
	We have implemented in $\texttt{SpaceMath v.2.0}$ LHC Higgs boson data (as well as projections for the HL-LHC and HE-LHC) and Lepton Flavor Violating processes (LFV). The former can be applied to any model that predicts corrections to the Higgs-fermions and Higgs-V ($V=W, Z$) couplings, while the LFV processes only can be studied for models with Flavor Changing Neutral Interactions at tree level. Specifically to models with both real and complex singlets, real and complex doublets and effective theories. We start describing the theoretical framework in which \texttt{SpaceMath v.2.0} works.
	\begin{enumerate}
		\item Higgs boson data
		\begin{enumerate}
			\item Signal strength modifiers $\mathcal{R}_X$ 
			\item Higgs boson coupling modifiers $\kappa_i$ 
		\end{enumerate}
	\item LFV processes
	\begin{enumerate}
		\item $h\to \ell_i \ell_j$ 
		\item Radiative processes $\ell_i\to\ell_j\gamma$,
		\item Muon anomalous magnetic dipole moment $a_{\mu}$,
		\item $\ell_i\to \ell_j \ell_j \ell_k$ decays.

				\end{enumerate}
	\end{enumerate} 
\subsection{LHC Higgs boson data}

\subsubsection{Signal strength modifiers $\mathcal{R}_{X}$}\label{muX}
For a production process $\sigma(pp\to H_i)$ and a decay $H_i\to X$, the signal strength is defined as follows:

\begin{equation}
	\mathcal{R}_{X}=\frac{\sigma(pp\to h)\cdot\mathcal{BR}(h\to X)}{\sigma(pp\to h^{\text{SM}})\cdot\mathcal{BR}(h^{\text{SM}}\to X)},
\end{equation}
where $\sigma(pp\to H_i)$ is the production cross section of $H_i$, with $H_i=h,\,h^{\text{SM}}$; here $h$ is the SM-like Higgs boson coming from an extension of the SM and $h^{\text{SM}}$ is the SM Higgs boson; $\mathcal{BR}(H_i\to X)$ is the branching ratio of the decay $H_i\to X$, with $X=b\bar{b},\;\tau^-\tau^+,\;\mu^-\mu^+,\;WW^*,\;ZZ^*,\;\gamma\gamma$.

In \texttt{SpaceMath v.2.0}, we consider the Higgs boson production cross section via the gluon fusion mechanism and we use the narrow width approximation:

\begin{equation}
	\mathcal{R}_{X}\approx\frac{\Gamma(h\to gg)\cdot\mathcal{BR}(h\to X)}{\Gamma(h^{\text{SM}}\to gg)\cdot\mathcal{BR}(h^{\text{SM}}\to X)}.
\end{equation}

\subsubsection{Higgs boson coupling modifiers $\kappa_i$}\label{kappaX}

The coupling modifiers $\kappa_i$ are introduced to quantify the deviations of the SM-like Higgs boson to other particles.
The coupling modifiers $\kappa_i$ for a production cross section or a decay mode, are defined as follows:

\begin{equation}\label{kappas}
\kappa_{pp}^2=\frac{\sigma(pp\to h)}{\sigma(pp\to h^{\text{SM}})}\;\text{or}\;\kappa_X^2=\frac{\Gamma(h\to X)}{\Gamma({h^\text{SM}}\to X)}.
\end{equation}

We consider tree-level Higgs boson couplings to different particles, i.e., $g_{hZZ^*}$, $g_{hWW^*}$, $g_{h\tau^-\tau^+}$, $g_{h\mu^-\mu^+}$, $g_{hb\bar{b}}$, as well as effective coupling modifiers $g_{hgg}$ and $g_{h\gamma\gamma}$ which describe gluon fusion production $\texttt{ggh}$ and the $h\to\gamma\gamma$ decay, respectively. 

\begin{table}

\caption{Current bounds of the observables implemented in $\texttt{SpaceMath v1.1}$}\label{Bounds}

\begin{centering}
\begin{tabular}{cc}
\hline 
Observable & Bound\cite{Workman:2022ynf}\tabularnewline
\hline 
\hline 
$\rm BR(h\to e\mu)$ & $<6.1\times10^{-5}$\tabularnewline
\hline 
$\rm BR(h\to e\tau)$ & $<2.2\times10^{-3}$\tabularnewline
\hline 
$\rm BR(h\to\mu\tau)$ & $<1.5\times10^{-3}$\tabularnewline
\hline 
$\rm BR(\mu\to e\gamma)$ & $<4.2\times10^{-13}$\tabularnewline
\hline 
$\rm BR(\tau\to e\gamma)$ & $<3.3\times10^{-8}$\tabularnewline
\hline 
$\rm BR(\tau\to\mu\gamma)$ & $<4.2\times10^{-8}$\tabularnewline
\hline 
$\rm BR(\tau\to3e)$ & $<2.7\times10^{-8}$\tabularnewline
\hline 
$\rm BR(\tau\to3\mu)$ & $<2.1\times10^{-8}$\tabularnewline
\hline 
$\rm BR(\mu\to3e)$ & $<1.0\times10^{-12}$\tabularnewline
\hline 
$\Delta a_{\mu}$ & $[1.92\times10^{-9},\,3.1\times10^{-9}]$\tabularnewline
\hline 
$\rm BR(B_{s}\to\mu\mu)$ & $[3.52\times10^{-9},\,3.8\times10^{-9}]$\tabularnewline
\hline 
$\rm BR(B_{d}\to\mu\mu)$ & $[0.98\times10^{-10},\,1.08\times10^{-10}]$\tabularnewline
\hline 
\end{tabular}
\par\end{centering}
\end{table}

	\subsection{Lepton Flavor Violating processes}
	\subsubsection{$h\to \ell_i \ell_j$ decays}\label{hlilj}
	The LFV processes $h\to \ell_i\ell_j$ ($\ell_{i,\,j}=\ell_{i,\,j}^-\ell_{i,\,j}^+$) where $\ell_i\ell_j=e\mu,\,e\tau,\,\tau\mu$ can arise at tree level in many models that extend to the SM. The relevant interactions can be extrated from the Yukawa Lagrangian
\begin{equation}
	\mathcal{L}_Y\supset -Y_{ij}\bar{\ell}_L^i \ell_R^jh+h.c.
\end{equation}	 
The correponding full decay width of the $h\to \ell_i \ell_j$ decays is given by:
\begin{equation}
\Gamma(h\to \bar{f}_i f_j)=\frac{N_c g^2_{h \bar{f}_i f_j }m_h}{128\pi}\Bigg[ 4-(\sqrt{\tau_{f_i}}+\sqrt{\tau_{f_j}})^2  \Bigg]^{3/2}\sqrt{4-(\sqrt{\tau_{f_i}}-\sqrt{\tau_{f_j}})^2},
\end{equation}
where $g_{h \bar{f}_i f_j}$ is the $h\bar{f}_i f_j$ coupling coming from an extesion of the SM, $Nc=3\,(1)$ is the color number for quarks (leptons), $m_h$ is the Higgs boson mass and $\tau_i=4m_i^2/m_h^2$. 

	\subsubsection{$\ell_i\to\ell_j\gamma$ decays}
	The effective Lagrangian for the $\ell_i\to\ell_j\gamma$\footnote{$\ell_i\to\ell_j\gamma$ stand for $\tau\to \mu \gamma, \tau\to e \gamma$ and $\mu\to e\gamma$} is given by
	\begin{equation}\label{EffLagrangian}
	\mathcal{L}_{\text{eff}}=C_L Q_{L\gamma} C_R Q_{R\gamma}+h.c.,
	\end{equation}
where the dim-5 electromagnetic penguin operators read
\begin{equation}\label{dim5Op}
Q_{L\gamma,\,R\gamma}=\frac{e}{8\pi^2}(\bar{\ell}_j\sigma^{\alpha\beta}P_{L,\,R}\ell_i)F_{\alpha\beta},
\end{equation}
where $F_{\alpha\beta}$ is the electromagnetic field strength tensor. The Wilson coefficients $C_{L,\,R}$ receive contributios at one-loop level and an important contribution from Barr-Zee two-loops level. For the particular case when $\ell_i=\tau$ and $\ell_j=\mu$, we assume the approximation $g_{\phi\mu\mu}\ll g_{\phi\tau\tau}$ and $m_\mu\ll m_{\tau}\ll m_\phi$. Under this assertion the one-loop Wilson coefficients $C_{L,\,R}$ simplify as follows \cite{Harnik:2012pb, Blankenburg:2012ex}
\begin{equation}\label{WilsonCoe1loop}
C_L^{1loop}\simeq \sum_{\phi}\frac{g_{\phi\tau\tau}g_{\phi\tau\mu}}{12m_\phi^2}\Bigg(-4+3\log\frac{m_\phi^2}{m_{\tau}^2}\Bigg),\;\;\;C_R^{2loop}\simeq\sum_{\phi}\frac{g_{\phi\tau\tau}g_{\phi\tau\mu}}{12m_\phi^2}\Bigg(-4+3\log\frac{m_\phi^2}{m_{\tau}^2}\Bigg),
\end{equation}
The numerical expressions for 2-loop contributions are given by
\begin{equation}\label{WilsonCoe2loop}
C_L^{2loop}=\sum_{\phi}g_{\phi\tau\mu}^*(-0.082g_{\phi tt}+0.11)/(m_\phi\text{GeV})^2.
\end{equation}

The rate for $\tau\to\mu\gamma$ is
\begin{equation}\label{ratetaumugamma}
\Gamma(\tau\to\mu\gamma)=\frac{\alpha m_{\tau}^2}{64\pi^4}(|C_L|^2+|C_R|^2).
\end{equation}
To obtain the correponding width decay of the processes $\mu\to e\gamma$ and $\tau\to e\gamma$, the replacements $\tau\to \mu,\, \mu\to e$ for the first decay and $\mu\to e$ for the second process from \eqref{EffLagrangian} to \eqref{ratetaumugamma} are required.

\subsubsection{Muon anomalous magnetic dipole moment ($\mu$AMDM)}

\begin{figure}[!htb]
		\centering
		\includegraphics[width=15cm]{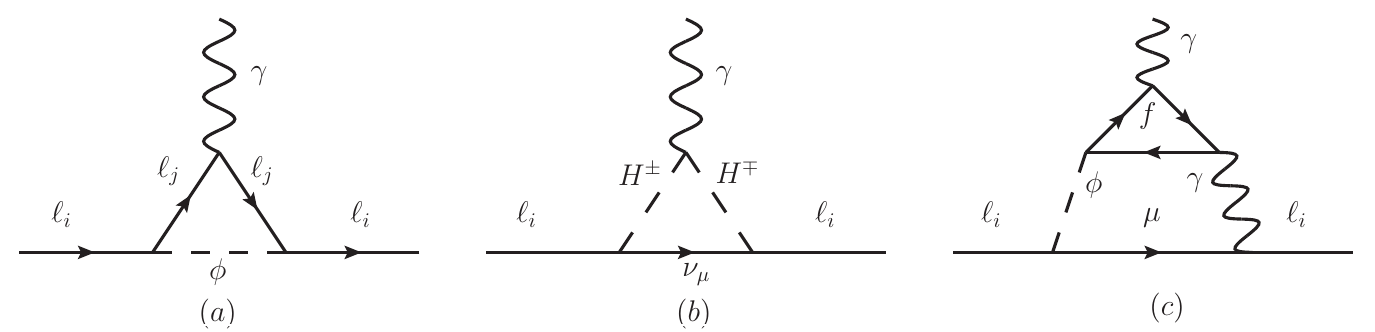}
		\caption{Feynman diagrams that contribute to $\mu$AMDM. Here $\phi$ represents a CP-even scalar, CP-odd scalar and the SM-like Higgs boson. $H^{\pm}$ stand for charged scalar bosons.}\label{FDmuonAMDM}
		\end{figure}

The Feynman diagrams that contribute to $\mu$AMDM are shown in Fig. \ref{FDmuonAMDM} 

The one-loop contribution in terms of Feynman parametritation method is given by
\begin{equation}
a_{\ell}^{a_1 a_2 a_3}=\underset{\underset{\ell'=2}{\ell=1,2,3}}{\sum}\frac{|\eta_{\ell\ell'}^{a_1a_2a_3}|^2m_{\ell}}{\sqrt{8}\pi^2}\int_0^1 dx\int_0^{1-x}dyG_k^{a_1a_2a_3}(x,\,y),
\end{equation} 
where the label $a_i$ stands for particles circulating inside the loop in each diagram in Fig. \ref{FDmuonAMDM}, the index $\ell\ell'$ denotes the entry of the Yukawa interaction. In the case of $\mu$AMDM $\ell'=2$, i.e., $\ell 2\to 12,\,22,\,32$. Notice that Eq. \ref{FDmuonAMDM} can be applied to any of the charged lepton. The term $\eta_{\ell\ell'}^{a_1 a_2 a_3}$ represents the coupling $a_1 a_2 a_3$.

The $G_{k=a}^{a_1a_2a_3}(x,\,y)$ function for the diagram (a) in Fig. \ref{FDmuonAMDM} reads
\begin{equation}
G_a^{\phi\ell_j\ell_j}(x,\,y)=(x+y)(m_{\ell_j}-m_{\ell_i}(x+y-1))/M_a^2,
\end{equation}
where $M_a^2=-m_{\phi}^2(x+y-1)+(x+y)(m_{\ell_j}^2+m_{\ell_i}^2(x+y-1))$, with $\ell_j=e,\,\mu,\,\tau$.

For diagram (b), $\eta_{\ell\ell'}^{H\pm H\mp \nu_i}=1$, while
\begin{equation}
G_b^{H\pm H\mp \nu_i}(x)=2m_{\ell_i}x/M_b^2,
\end{equation}
with $M_b^2=(m_{\ell_i}^2 x-m_{H^{\pm}}^2)$. There are also Barr-Zee two-loop contributions to the $\mu$AMDM. The dominant contribution is given by 
\begin{equation}
a_{\ell_i}^{\rm two-loop}=\frac{\alpha^2}{8\pi^2s_W^2}\frac{m_{\ell_i}^2 g_{A\ell_i\bar{\ell}_i}}{m_W^2}\sum_{f=t,\,\tau,\,b}N_c^f Q_f^2 r_f f(r_f) g_{A\ell_i\bar{\ell}_i},
\end{equation}
where $r_f=(m_f/m_A)$, $m_f$ is the fermion mass, $N_c^f=1(3)$ for leptons (quarks), $Q_f$ is the electric charge of fermions and $g_{Aff}$ is given by the following Lagrangian 
\begin{equation}
\mathcal{L}=i\frac{g m_f g_{Aff} }{2m_W}\bar{f}\gamma^5 f A,
\end{equation}
and finally
\begin{equation}
f(x)=\int_0^1 \frac{\log(\frac{x}{y(1-y)})}{x-y(1-x)}dy.
\end{equation}

	\section{Installation and first steps}\label{SeccionIII}
	\subsection{Installation}

	Run the following instructions in a Notebook of \texttt{Mathematica}
	
	\begin{center}\resizebox{\textwidth}{!}{%
		\fbox{\begin{minipage}{\dimexpr\textwidth+25\fboxsep-2\fboxrule\relax}	
				\centering\texttt{{
				Import["https://raw.githubusercontent.com/spacemathapp/SpaceMath-v.2.0/alpha/Install.m"]}}	\[ 
				{\texttt{InstallSpaceMath[]}}
				\] 
		\end{minipage}}}
	\end{center}

	Note that an error may appear because the quotation marks (""); this can be resolved by deleting and then explicitly writing both quotation marks.

		To delete \texttt{SpaceMath} automatically, the user only has to execute the following instruction:
 
 	\begin{center}\resizebox{\textwidth}{!}{%
 			\fbox{\begin{minipage}{\dimexpr\textwidth+25\fboxsep-2\fboxrule\relax}	
 					\centering\texttt{{
 							DeleteSpaceMath[]}}
 		\end{minipage}}}
 	\end{center}

	\subsection{First steps}
	
	Firts of all, we define in Table \ref{arguments} the arguments that are commons in the most commands described below. Thus, we encourage the reader to become familiar with them. 

\begin{table}[!htb]
\caption{Description of arguments shared in all \texttt{SpaceMath v.2.0} commands.}\label{arguments}

\begin{centering}
\begin{tabular}{|c|c|}
\hline 
Argument & Description\tabularnewline
\hline 
\hline 
$\texttt{xi, (i=1, 2, 3, 4)}$ & Parameters to constraint\tabularnewline
\hline 
$\texttt{ximin (\texttt{ximax})}$ & Initial (final) value of the interval to evaluate\tabularnewline
\hline 
$\texttt{xilabel}$ & Label the column \texttt{i=1, 2, 3, 4} to be plotted\tabularnewline
\hline 
$\texttt{NN}$ & Random values to generate\tabularnewline
\hline
$\texttt{ghXX}$ & Represents the $g_{hXX}$ coupling, where \texttt{XX}$=b\bar{b},t\bar{t}, \tau^-\tau^+,\mu^-\mu^+, WW, ZZ$ \tabularnewline
\hline 
\end{tabular}
\par\end{centering}
\end{table}


To generate random points in accordance with experimental measurements, we use the following instructions.
\subsection*{Signal strengths}
\subsubsection*{\underline{Signal strength $\mathcal{R}_b$}}
	
\begin{eqnarray}\label{NewRbRandom}	
	\texttt{Rb[ghtt,ghbb,x1,x1min,x1max,x1label,x2,x2min,x2max,x2label,}&\\ \nonumber
	\texttt{x3,x3min,x3max,x3label,x4,x4min,x4max,x4label,NN]}
\end{eqnarray}
\subsubsection*{\underline{Signal strength $\mathcal{R}_\tau$}}
\begin{eqnarray}\label{NewRtauRandom}
	\texttt{Rtau[ghtt,ghbb,ghtautau,x1,x1min,x1max,x1label,x2,x2min,x2max,}&\\ \nonumber
	\texttt{x2label,x3,x3min,x3max,x3label,x4,x4min,x4max,x4label,NN]}
\end{eqnarray}
\subsubsection*{\underline{Signal strength $\mathcal{R}_V$, ($V=Z,\,W$)}}
\begin{eqnarray}\label{NewRVRandom}
	\texttt{RV[ghtt,ghbb,ghVV,x1,x1min,x1max,x1label,x2,x2min,x2max,}&\\ \nonumber
	\texttt{x2label,x3,x3min,x3max,x3label,x4,x4min,x4max,x4label,NN]}
\end{eqnarray}
\subsubsection*{\underline{Signal strength $\mathcal{R}_{\gamma}$}}
\begin{eqnarray}\label{NewRgamRandom}
	\texttt{Rgam[ghtt,ghbb,ghWW,gCH,mCH,x1,x1min,x1max,x1label,x2,x2min, x2max,}&\\ \nonumber
	\texttt{x2label,x3,x3min,x3max,x3label,x4,x4min,x4max,x4label,NN]}
\end{eqnarray}
\subsubsection*{\underline{All the signal strengths $\mathcal{R}_X$}}
\begin{eqnarray}\label{NewRALLRandom}
	\texttt{RALL[ghtt,ghbb,ghZZ,ghWW,gCH,mCH,x1,x1min,x1max,x1label,x2,x2min, }&\\ \nonumber
	\texttt{x2max,x2label,x3,x3min,x3max,x3label,x4,x4min,x4max,x4label,NN]}
\end{eqnarray}
\subsubsection*{\underline{Intersection of all the signal strengths $\mathcal{R}_X$}}
\begin{eqnarray}\label{NewRintersectionRandom}
	\texttt{Rintersection[ghtt,ghbb,ghZZ,ghWW,ghtautau,gCH,mCH,}&\\ \nonumber
	\texttt{x1,x1min, x1max,\textit{x1label},x2,x2min,x2max,\textit{x2label},x3,x3min,}&\\ \nonumber
	\texttt{x3max,\textit{x3label},x4,x4min,x4max,\textit{x4label},NN]}
\end{eqnarray}
 Here, the \texttt{RALL} command include all the $\mathcal{R}_X$'s to be plotted in the same plot while the instruction \texttt{Rintersection} generates random points that satisfy all the $\mathcal{R}_X$'s.
In Eqs. \eqref{NewRgamRandom}, \eqref{NewRALLRandom} and \eqref{NewRintersectionRandom}, the arguments \texttt{gCH} and \texttt{mCH} stand for the $g_{hH^+H^-}$ coupling and the mass of a charged scalar boson, respectively. All the points that meet the experimental restrictions will be exported  to \texttt{\$UserDocumentDirectory}\footnote{You can execute this command in a notebook of \texttt{Mathematica} and will show the location path.} (Documents).

\texttt{SpaceMath v.2.0} has its own command to graph the $\mathcal{R}_X$'s. After random points generation, it can accomplished with the following instruction: 
\begin{equation}\label{PlotRbRandomSelectAxis}
	\texttt{PlotRX[\textit{x1label}, \textit{x2label}, \textit{x3label}, \textit{x4label}]}.
\end{equation}
The user must make the replacement \texttt{X $\to$ b, tau, W, Z, gam, ALL, intersection} to generate the correponding graph for $\mathcal{R}_b$, $\mathcal{R}_\tau$, $\mathcal{R}_W$, $\mathcal{R}_Z$, $\mathcal{R}_\gamma$, $\mathcal{R}_{\text{ALL}}$, $\mathcal{R}_{\text{intersection}}$, respectively.

Once the main commands have been defined, we focus on the particular case of  $\mathcal{R}_{\text{ALL}}$, which generates points including all the $\mathcal{R}_{\text{X}}$'s. For this purpose, we consider the Yukawa interaction Lagrangian of the Two-Higgs Doublet Model of type III \cite{Branco:2011iw, Botella:2009pq, Botella:2015hoa, Arroyo:2013tna, Cruz:2019vuo, HernandezSanchez:2012eg, BarradasGuevara:2010xs, Arroyo-Urena:2015uoa, GomezBock:2005hc, HernandezSanchez:2010zz, Arroyo-Urena:2019qhl, Arroyo-Urena:2020mgg}, which is given by

	\begin{eqnarray}\label{YukLagrangian}
	\mathcal{L}_{Y}^{\text{THDM-III}} & = & \frac{g}{2}\left(\frac{m_{\ell}}{m_{W}}\right)\bar{\ell}_{i}\left[\frac{\cos\alpha}{\cos\beta}\delta_{ij}+\frac{\sqrt{2}\sin(\alpha-\beta)}{g\cos\beta}\left(\frac{m_{W}}{m_{\ell}}\right)\left(\frac{\sqrt{m_i m_j}}{v}\chi_{ij}\right)\right]\ell_{j}H\nonumber \\
		& + & \frac{g}{2}\left(\frac{m_{\ell}}{m_{W}}\right)\bar{\ell}_{i}\left[-\frac{\sin\alpha}{\cos\beta}\delta_{ij}+\frac{\sqrt{2}\cos(\alpha-\beta)}{g\cos\beta}\left(\frac{m_{W}}{m_{\ell}}\right)\left(\frac{\sqrt{m_i m_j}}{v}\chi_{ij}\right)\right]\ell_{j}h\nonumber \\
		& + & i\frac{g}{2}\left(\frac{m_{\ell}}{m_{W}}\right)\bar{\ell}_{i}\left[\tan\beta\delta_{ij}+\frac{\sqrt{2}}{g\cos\beta}\left(\frac{m_{W}}{m_{\ell}}\right)\left(\frac{\sqrt{m_i m_j}}{v}\chi_{ij}\right)\right]\gamma^{5}\ell_{j}A\nonumber \\
		& + & \frac{g}{2}\left(\frac{m_{u}}{m_{W}}\right)\bar{u}_{i}\left[\frac{\sin\alpha}{\sin\beta}\delta_{ij}-\frac{\sqrt{2}\sin(\alpha-\beta)}{g\sin\beta}\left(\frac{m_{W}}{m_{u}}\right)\left(\frac{\sqrt{m_i m_j}}{v}\chi_{ij}\right)\right]u_{j}H\\
		& + & \frac{g}{2}\left(\frac{m_{u}}{m_{W}}\right)\bar{u}_{i}\left[\frac{\cos\alpha}{\sin\beta}\delta_{ij}-\frac{\sqrt{2}\cos(\alpha-\beta)}{g\sin\beta}\left(\frac{m_{W}}{m_{u}}\right)\left(\frac{\sqrt{m_i m_j}}{v}\chi_{ij}\right)\right]u_{j}h\nonumber \\
		&+&i\frac{g}{2}\left(\frac{m_{u}}{m_{W}}\right)\bar{u}_{i}\left[\cot\beta\delta_{ij}+\frac{\sqrt{2}}{g\sin\beta}\left(\frac{m_{W}}{m_{u}}\right)\left(\frac{\sqrt{m_i m_j}}{v}\chi_{ij}\right)\right]\gamma^{5}u_{j}A,\nonumber
		\end{eqnarray}
where $i$ and $j$ stand for the fermion flavors, with $i\neq j$, in general. As far as the type-down quark interactions, it is similar to lepton part with the exchange $\ell\to d$ and $m_\ell\to m_{d}$. In addition to the SM-like Higgs boson, represented by $h$, the THDM-III predicts two neutral spin-0 particles denoted by $H$ and $A$ in Eq. \eqref{YukLagrangian}\footnote{The Yukawa Lagrangian in Eq. \ref{YukLagrangian} only shows the neutral interactions, but the model also predicts two charged scalars, no included there.} \\

\begin{figure}[!htb]
	\centering
	\includegraphics[width=8cm]{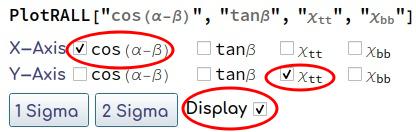}
	\caption{In this menu the user select the plane to be plotted.}\label{menuRall}
\end{figure}

\begin{figure}[!htb]
 	\centering
 	\includegraphics[width=8cm]{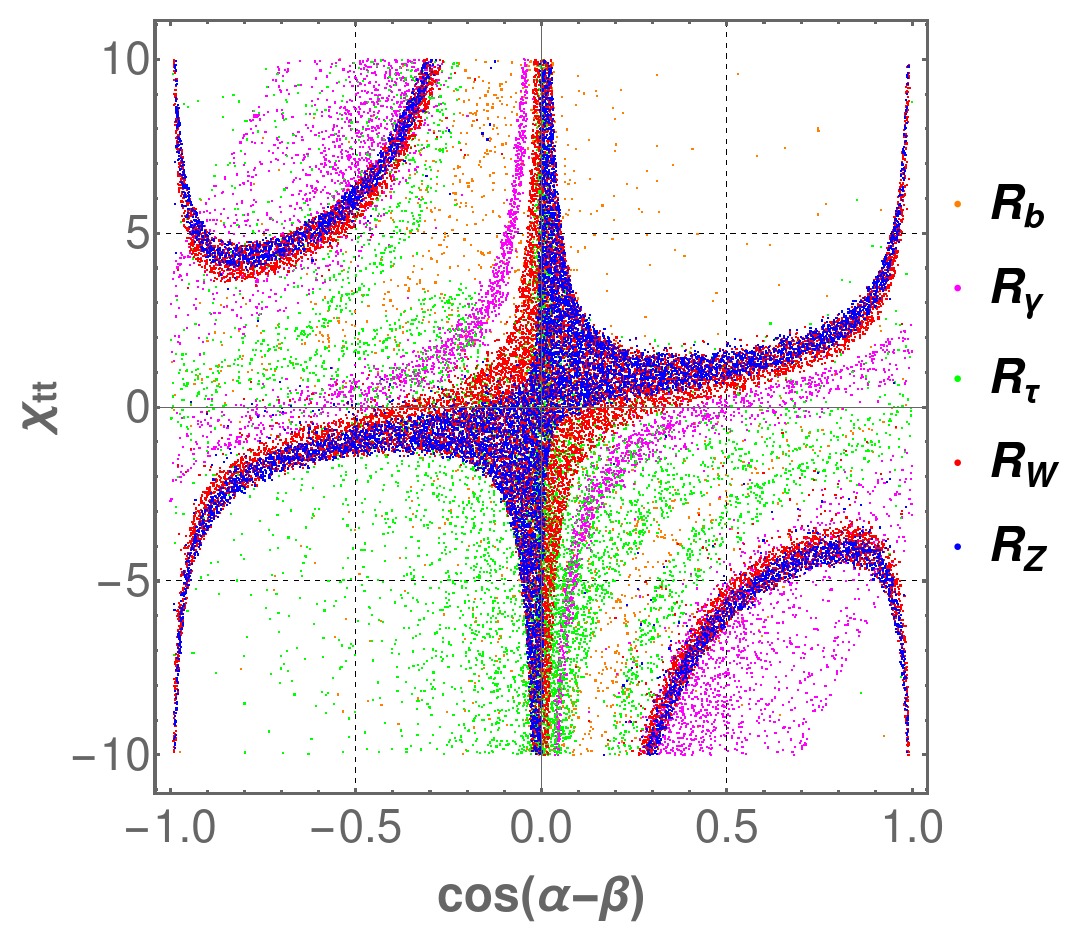}
 	\caption{Plot generated by \texttt{SpaceMath v.2.0} via the command \texttt{PlotRALL}. Points orange (purple, green, red and blue) are the ones that satisfy $\mathcal{R}_b$ ($\mathcal{R}_{\gamma}$, $\mathcal{R}_\tau$, $\mathcal{R}_W$ and $\mathcal{R}_Z$).}\label{RALLplot}
 \end{figure}

The explicit steps to follow read:
\begin{enumerate}
	\item Open a notebook of \texttt{Mathematica} and load \texttt{SpaceMath v.2.0} by typing \texttt{<<SpaceMath`},
	\item Define the couplings as a function of the parameters to be constrained. In the theoretical framework of THDM-III (Eq. \ref{YukLagrangian}), it is given by: 
		\begin{itemize}
				\item \texttt{ghtt[a$\_$,chitt$\_$,Cab$\_$,tb$\_$]:=(g/2) (mt/mW)((Cos[a]/(tb*Cos[ArcTan[tb]]))}\\ 
			\texttt{-(Sqrt[2] Cab/(g*tb*Cos[ArcTan[tb]]) (mW/mt)*(mt/vev*chitt)))}
			
			\item \texttt{ghbb[a$\_$,chibb$\_$,Cab$\_$,tb$\_$]:=(g/2) (mb/mW) (((-Sin[a]tb)/Sin[ArcTan[tb]])}\\
			\texttt{+(Sqrt[2] (Cab*tb)/(g*Sin[ArcTan[tb]]) (mW/mb)(mb/vev*chibb)))}
			
			\item \texttt{ghtautau[a$\_$,chitata$\_$,Cab$\_$,tb$\_$]:=(g/2)(mtau/mW)(((-Sin[a]tb)/Sin[ArcTan[tb]])}\\
			\texttt{+(Sqrt[2] (Cab*tb)/(g*Sin[ArcTan[tb]])(mW/mtau)(mtau/vev*chitata)))}
			
			\item \texttt{gCH[tb$\_$, Cab$\_$] := mW*Sin[ArcTan[tb]-(ArcCos[Cab]+ArcTan[tb])]+}\\
				\texttt{mZ/(2 CW)Cos[2 ArcTan[tb]]Sin[ArcTan[tb]+(ArcCos[Cab]+ ArcTan[tb])]}
			
			\item \texttt{ghWW[sab$\_$]:=gw*mW*sab}
			\item \texttt{ghZZ[sab$\_$]:=gz*mZ*sab}
	\end{itemize}
where \texttt{a, Cab, tb, chitt(bb) \text{and} sab} are identified with $\alpha,\,\cos(\alpha-\beta),\,\tan\beta,\,\chi_{tt(bb)},\,\sin(\alpha-\beta)$, respectively, in Eq. \eqref{YukLagrangian},
\item Later, we execute the instruction\\

\texttt{RALL[ \\
	ghtt[ArcCos[Cab]+ArcTan[tb],chitt,Cab,tb], }\\
\texttt{ghbb[ArcCos[Cab]+ArcTan[tb],chibb,Cab,tb],}\\
\texttt{ghZZ[Sqrt[$\texttt{1-Cab}^2$]],}\\
\texttt{ghWW[Sqrt[$\texttt{1-Cab}^2$]],}\\
\texttt{ghtautau[ArcCos[Cab]+ArcTan[tb],1,Cab,tb],}\\
\texttt{gCH[tb,Cab],600,Cab,-1,1,"$\cos(\alpha-\beta)$",tb,0.1,50,"$\tan\beta$",}\\
\texttt{chitt,-10,10,"$\chi_{tt}$",Abb,-10,10,"$\chi_{bb}$",100000}\\
\texttt{];}
\\
Notice that we have made the following definition: \texttt{a=ArcCos[Cab]+ArcTan[tb]}
\item Once the random points have been generated (point 3), the command to plot them is the following 
\begin{itemize}
	\item \texttt{PlotRALL["$\cos(\alpha-\beta)$","$\tan\beta$", "$\chi_{tt}$", "$\chi_{bb}$"],}\label{xiLables}
\end{itemize}\label{PlotALL}
where \texttt{\textit{x1label}}=\texttt{"}$\cos(\alpha-\beta)$\texttt{"}, \texttt{\textit{x2label}}=\texttt{"}$\tan\beta$\texttt{"}, \texttt{\textit{x3label}}=\texttt{"}$\chi_{tt}$\texttt{"} and  \texttt{\textit{x3label}}=\texttt{"}$\chi_{bb}$\texttt{"}]. Notice that in Eq. \eqref{PlotALL} the arguments are enclosed in quotes (\texttt{"..."}).
	\end{enumerate}
After executing the instruction \texttt{PlotRALL[...]}, a menu will be displayed.  If the user has selected the checkboxes as shown in Fig. \ref{menuRall}, then the Fig. \ref{RALLplot} will be generated.  

\subsection*{Machine Learning}
We have implemented algoritms of machine learning in \texttt{SpaceMath v.2.0} to generate specific Benchmark Points useful to evaluate the calculations of physical observables of interest. The algorithms included in \texttt{SpaceMath v.2.0} are \textit{Linear Regression, Decision Trees, Gaussian Process, Gradient Boosted Trees and Neural Networks}. Once the user has generated the random points through the instructions \texttt{RX}, the command that involke these algorithms is as follows.

\begin{eqnarray}\label{MLalgoritms}
	\texttt{Ralgorithm[x1label,x2label,x3label,x4label]},
\end{eqnarray}
where \texttt{algorithm $\to$ LinearRegression, DecisionTrees, GaussianProcess,\\ GradientBoostedTrees, NeuralNetworks}.

\begin{figure}[!htb]
	\centering
	\includegraphics[width=8cm]{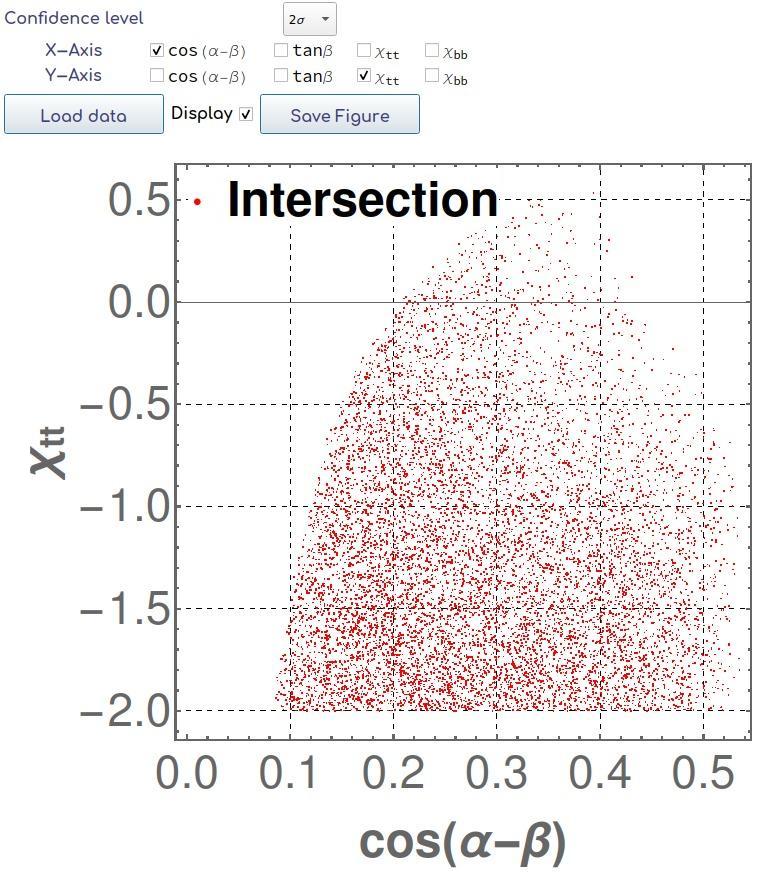}
	\caption{Plot generated by \texttt{SpaceMath v.2.0} via the command \texttt{PlotRintersection}.}\label{RplotInter}
\end{figure}

We suggest using \texttt{Rintersection} as this considers the points that pass the test of all $\mathcal{R}_X$'s; the user also can use any \texttt{RX} of interest, though. In this way, in order to illustrate how \texttt{SpaceMath} with Machine Learning works, we consider \texttt{Rintersection} applied to the THDM-III:

\begin{itemize}
	\item \texttt{Rintersection[}\\
\texttt{ ghtt[ArcCos[Cab]+ArcTan[tb],chitt,Cab,tb],}\\ 
\texttt{ ghbb[ArcCos[Cab]+ArcTan[tb],chibb,Cab,tb],}\\
\texttt{ ghZZ[Sqrt[$\texttt{1-Cab}^2$]],}\\
\texttt{ ghWW[Sqrt[$\texttt{1-Cab}^2$]],}\\
\texttt{ ghtautau[ArcCos[Cab]+ArcTan[tb],1,Cab,tb],}\\
\texttt{ gCH[tb,Cab],}\\
\texttt{ 600,}\\
\texttt{ Cab,-1,1,"$\cos(\alpha-\beta)$",}\\
\texttt{ tb,0.1,50,"$\tan\beta$",}\\
\texttt{ chitt,-2,1,"$\chi_{tt}$",}\\
\texttt{ chibb,-1,2,"$\chi_{bb}$",}\\
 \texttt{500000000}\\
\texttt {].}
\end{itemize}
To plot the points generated via the command \texttt{Rintersection[...]} the user can use the instruction:

\begin{itemize}
	\item \texttt{PlotRintersection["$\cos(\alpha-\beta)$","$\tan\beta$", "$\chi_{tt}$", "$\chi_{bb}$"],}
\end{itemize}
whose output will be a graph as shown in Fig \eqref{RplotInter}
Meanwhile, the instruction to generate the Benchmark Points by the different Machine Learning methods are the following:
\begin{itemize}
	\item \texttt{RLinearRegression["$\cos(\alpha-\beta)$","$\tan\beta$", "$\chi_{tt}$", "$\chi_{bb}$"],}
	\item \texttt{RDecisionTrees["$\cos(\alpha-\beta)$","$\tan\beta$", "$\chi_{tt}$", "$\chi_{bb}$"],}
	\item \texttt{RGradientBoostedTrees["$\cos(\alpha-\beta)$","$\tan\beta$", "$\chi_{tt}$", "$\chi_{bb}$"],}
	\item \texttt{RNeuralNetworks["$\cos(\alpha-\beta)$","$\tan\beta$", "$\chi_{tt}$", "$\chi_{bb}$"],}
	\item \texttt{RGaussianProcess["$\cos(\alpha-\beta)$","$\tan\beta$", "$\chi_{tt}$", "$\chi_{bb}$"].}
\end{itemize}
The Benchmarck Points found are given in Table \ref{BMPsML}.  
\begin{table}[!htb]
	
	\caption{Benchmark Points predicted by the Machine Learning. algorithms}\label{BMPsML}
	
	\begin{centering}
		\begin{tabular}{|c|c|c|c|c|}
			\hline 
			Method & $\cos(\alpha-\beta)$ & $\tan\beta$ & $\chi_{tt}$ & $\chi_{bb}$\tabularnewline
			\hline 
			\hline 
			Linear Regression & $0.2865$ & $0.2619$ & $-1.2418$ & $0.7503$\tabularnewline
			\hline 
			Decision Trees & $0.2858$ & $0.2610$ & $-1.2473$ & $0.7630$\tabularnewline
			\hline 
			Gradiant Boosted Trees  & $0.2858$ & $0.2642$ & $-1.2675$ & $0.7547$\tabularnewline
			\hline 
			Neural Networks & $0.2820$ & $0.2640$  & $-1.1226$ &$0.7637$ \tabularnewline
			\hline 
		\end{tabular}
		\par\end{centering}
\end{table}

\subsection*{Lepton Flavor Violating processes}

\subsubsection*{\underline{$h\to \ell_i \ell_j$ decay}}
The command to generate the parameter space allowed by the experimental measurement on the BR($h\to\ell_i\ell_j$) is the following
\begin{eqnarray}\label{hlilj}
&&\texttt{hlilj[ghlilj\_,\,x1\_,\,x1min\_,\,x1max\_,\,x1label\_,\,x2\_,\,x2min\_,\,x2max\_,\,x2label\_}\nonumber \\ 
&& \texttt{x3\_,\,x3min\_,\,x3max\_,\,x3label\_,\,x4\_,\,x4min\_,\,x4max\_,\,x4label\_,\,NN\_]},
\end{eqnarray}
where $\texttt{ghlilj}$ stands for the $h\ell_i\ell_j$ coupling, while the rest of the parameters are defined in Table \ref{arguments}.
 The command \texttt{hlilj} in Eq. \eqref{hlilj} exports an output file with values that agree with upper limits on $BR(h\to \ell_i \ell_j)$, its name is labeled as \texttt{hlilj.csv} and it will be saved in $\texttt{\$UserDocumentsDirectory}$. The command to graph the data generated by the command in Eq. \eqref{hlilj} is given by
\begin{eqnarray}\label{PlothTauMu}
\texttt{Plothlilj[x1label\_, x2label\_, x3label\_, x4label\_]}
\end{eqnarray}

Assuming the interactions shown in Eq.\eqref{YukLagrangian}, the $\texttt{SpaceMath}$ code (when $\ell_i=\tau$ and $\ell_j=\mu$) is given by 
\begin{eqnarray}
&&\texttt{hTauMu[ghtaumu[chitaumu, Cab, tb], Cab, -1, 1, "Cab", tb,0.1, 50,"tb"}\nonumber \\ 
&&\texttt{chitaumu, 0, 1,"chitaumu", x4, 0, 0,"x4label", 10000]},
\end{eqnarray} 
where 
\begin{equation}
\texttt{ghtaumu[chitaumu, Cab, tb]=$\frac{\cos(\alpha-\beta)}{\sqrt{2}\cos\beta}\frac{\sqrt{m_{\tau}m_{\mu}}}{v}\chi_{\tau\mu}=\frac{\cos(\alpha-\beta)\tan\beta}{\sqrt{2}\sin\beta}\frac{\sqrt{m_{\tau}m_{\mu}}}{v}\chi_{\tau\mu}$}
\end{equation}
is the $h\tau\mu$ coupling that depends on the parameters to be constrained. Note that the $h\tau\mu$ coupling depends only on three parameters, namely, \texttt{x1=Cab}, \texttt{x2=tb}, \texttt{x3=chitaumu}. In this case, the fourth parameter \texttt{x4} is free, so it is recommended to set \texttt{x4min=0} and \texttt{x4max=0}. 
Again, to plot the data we use the command in Eq. \eqref{PlothTauMu}.
\begin{equation}\label{PlotTauMu}
\texttt{PlothTauMu[\text{"}$\cos(\alpha-\beta)$\text{"}, \text{"}$\tan\beta$\text{"}, \text{"}$\chi_{\tau\mu}$\text{"}, \text{" "}]},
\end{equation} 
Notice that the instruction \eqref{PlothTauMu} works as the command \eqref{PlotALL}.

\begin{figure}[!htb]
		\centering
		\includegraphics[width=8cm]{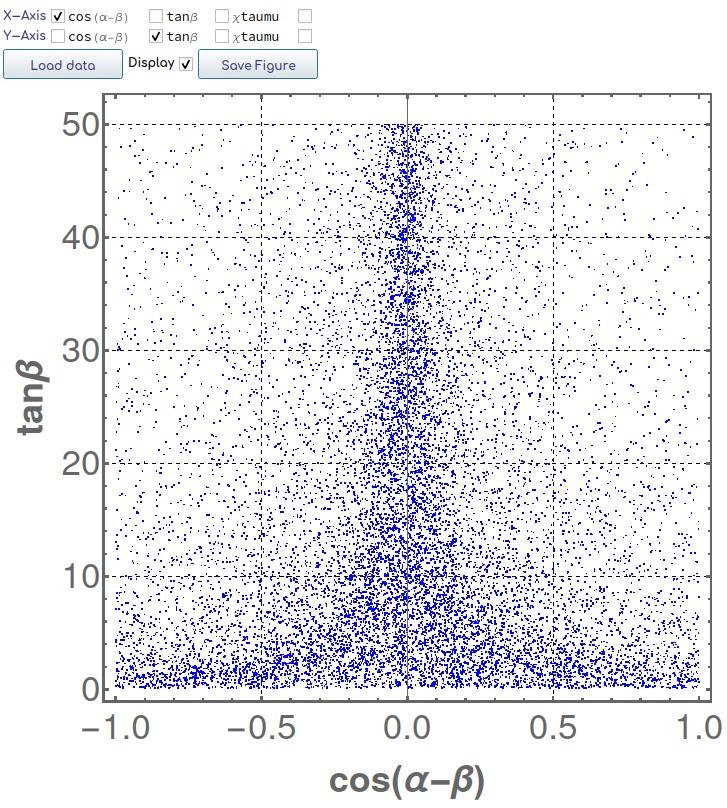}
		\caption{Graph generated by \texttt{SpaceMath v.2.0} via the instruction in Eq. \eqref{PlotTauMu}.}\label{htaumuFig}
		\end{figure}

Figure \ref{htaumuFig} shows the plot generated by executing the command in Eq.  \eqref{PlotTauMu}.

\subsubsection*{\underline{$\ell_i \to \ell_j\gamma$ decays}}
As far as the $\ell_i\to \ell_j\gamma$ decays are concerned, the command to generate the parameter space allowed by current upper bounds on $\rm BR(\tau\to\mu\gamma)$ (see Table \ref{Bounds}) reads
\begin{eqnarray}\label{liljgamma}
&&\texttt{TauMuGamma[ghtaumu\_,\,ghtautau\_,\,gAtaumu\_,\,gAtautau\_,\,gHtaumu\_,\,gHtautau\_,\,}\nonumber\\
&&\texttt{ghtt\_,\,gHtt\_,\,gAtt\_,\,mH\_,\,mA\_,\,x1\_,\, x1min\_,\,x1max\_,\,x1label\_,\,x2\_,\,x2min\_,\,x2max\_,\,} \nonumber \\
&&\texttt{x2label\_,\,x3\_,\,x3min\_,\,x3max\_,\,x3label\_,\,x4\_,\,x4min\_,\,x4max\_,\,x4label\_,\,NN\_],} 
\end{eqnarray}
where \texttt{PHItaumu}, \texttt{PHItautau}, \texttt{PHItt} are the $\Phi\tau\mu$, $\Phi\tau\tau$ and $\Phi tt$ couplings, respectively. The command in Eq. \eqref{liljgamma} exports an output file (\texttt{TauMuGamma.csv}) to $\texttt{\$UserDocumentsDirectory}$ with values in accordance with the upper bounds on $\rm BR(\ell_i\to\ell_j\gamma)$ (see Table \ref{Bounds}). 

To analyze the model parameter space via $\mu\to e\gamma$  users must make replacements
\begin{eqnarray}
&&\texttt{TauMuGamma}\to \texttt{MuEGamma}, \nonumber\\
&&\texttt{gPHItaumu}\to \texttt{gPHImue},\,\,\, \texttt{gPHItautau}\to \texttt{gPHImumu},\nonumber
\end{eqnarray}
where \texttt{PHI = h, H, A}.
And analogously for the $\tau\to e\gamma$ decay
\begin{eqnarray}
&&\texttt{TauMuGamma}\to \texttt{TauEGamma}, \nonumber\\
&&\texttt{gPHItaumu}\to \texttt{gPHItaue}.\nonumber
\end{eqnarray}


To generate the corresponding plot of the parameters space we use 
\begin{eqnarray}
\texttt{PlotLiLjGamma[x1label, x2label, x3label, x4label]}.
\end{eqnarray}
For the particular case when $\texttt{Li}=\tau$ and $\texttt{Lj}=\mu$, the specific instruction to follow is
\begin{eqnarray}
\texttt{PlotTauMuGamma[\text{"}$\chi_{\tau\mu}$\text{"},\text{"}$\chi_{\tau\tau}$\text{"}, \text{" "}, \text{" "}  ]}.
\end{eqnarray}


The procedure for analyzing the observables $\ell_i\to \ell_j \ell_k \bar{\ell}_k$ and $\delta a_\mu$ is similar to the previous instructions. User can follow the path in \ref{path} to see preloaded examples in \texttt{SpaceMath v.2.0}.
\begin{eqnarray}\label{path}
&\texttt{Tau3Mu[...]$\to$SpaceMath-2.0 $\to$ Observables $\to$ LFV $\to$ Tau3Mu},\nonumber\\
&\texttt{muonAMDM[...]$\to$SpaceMath-2.0 $\to$ Observables $\to$ LFV $\to$ muonAMDM}.
\end{eqnarray}

	
	\section{Validation}\label{SeccionIV}
	In order to validate $\texttt{SpaceMath v.2.0}$, we apply the coupling modifiers $\kappa_i$ defined in eq. \eqref{kappas} to the Two-Higgs Doublet Model of Type I and II (THDM-I, II). In Ref. \cite{Craig:2012vn} are reported $\kappa_b$ and $\kappa_V$ in the context of these models. To reproduce these results via \texttt{SpaceMath v.2.0} the only we need to know are the model couplings, which are given in Table \ref{AcoplosTHDM}. The commands to evaluate $\kappa_b$ and $\kappa_V$ are displayed in Table \ref{commadKbkV}.
	
	\begin{table}
		\caption{First column: THDM-I, -II couplings. Second column: coupling defined in \texttt{SpaceMath} \texttt{(v(V)=,\,z(Z),\,w(W))}. Third column: \texttt{SpaceMath} code.}\label{commadKbkV}
		\centering{}{\tiny{}}%
		\begin{tabular}{|c|c|c|}
			\hline 
			{\tiny{}Coupling} & {\tiny{}Input to SpaceMath} & {\tiny{} Command $\kappa_i$ 
			}\tabularnewline
			\hline 
			\hline 
			{\tiny{}$g_{hbb}^{THDM-I}=\frac{gm_{b}}{2m_{W}}\left(\frac{\cos\alpha}{\sin\beta}\right)$} & {\tiny{}$\texttt{ghbb[Sa\_,Tb\_,Cb\_]:=g*mb*Sqrt[1-Sa\textasciicircum2]/(2*mW*Tb*Cb)}$} & {\tiny{}$\texttt{kb[\texttt{ghbb[Sa,Tb,Cos[ArcTan[Tb]]}]]}$}\tabularnewline
			\hline 
			{\tiny{}$g_{hbb}^{THDM-II}=\frac{gm_{b}}{2m_{W}}\left(\frac{-\sin\alpha}{\cos\beta}\right)$} & {\tiny{}$\texttt{ghbb[Sa\_,Tb\_,Sb\_]:=-g*mb*Sa*Tb/(2*mW*Sb) }$} & {\tiny{}$\texttt{kb[\texttt{ghbb[Sa,Tb,Sin[ArcTan[Tb]]}]]}$}\tabularnewline
			\hline 
			{\tiny{}$g_{hVV}^{THDM-I,-II}=g_{V}m_{V}\sin(\beta-\alpha)$} & {\tiny{ $\textrm{ghVV[Tb\_,Cb\_,Sb\_,Sa\_]:=((Tb*Cb*Sqrt[1-Sa\textasciicircum2])-(Sb/Tb*Sa))*(gv*mV)}$} }& {\tiny{}$\texttt{kV[ghVV[Tb, Cos[ArcTan[Tb]], Sin[ArcTan[Tb]], Sa]]}$} \tabularnewline
			\hline 
		\end{tabular}{\tiny\par}
	\end{table}
	
\begin{figure}[!h] 
		\centering
		\includegraphics[width=5cm]{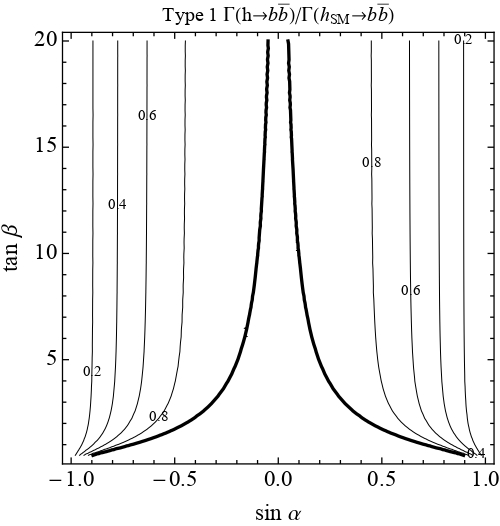} 
		\includegraphics[width=6.5cm]{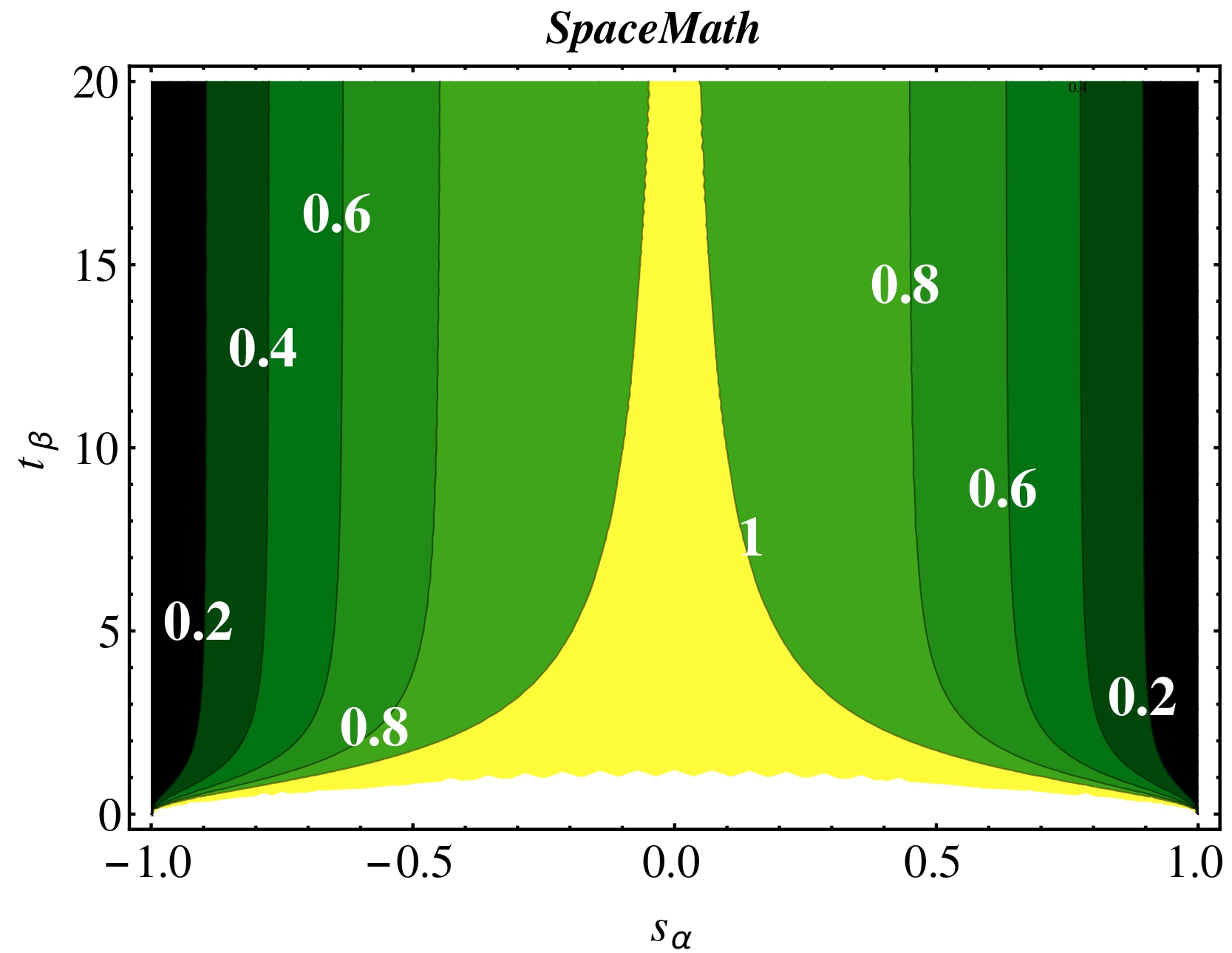}
		\caption{Contours of $\Gamma(h \to b \bar b) / \Gamma(h_{SM} \to b \bar b)$ 
			for the SM-like Higgs boson as a function of $\sin \alpha$ and 
			$\tan \beta$ in Type 1 THDM. Left: figure taken from \cite{Craig:2012vn} and Right: figure generated by \texttt{SpaceMath v.2.0}. 
		}
		\label{fig:widthbb1}
	\end{figure}

\begin{figure}[!h] 
		\centering
		\includegraphics[width=5cm]{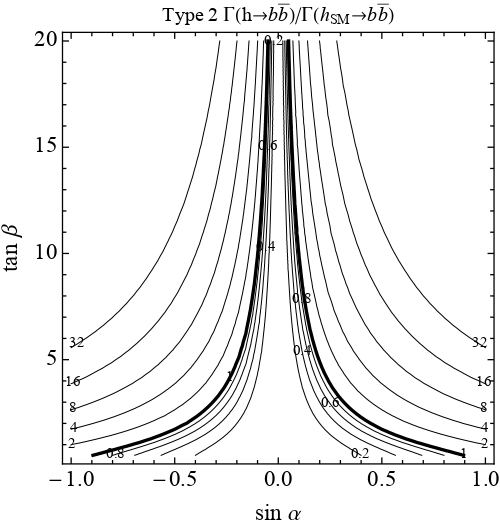} 
		\includegraphics[width=6.5cm]{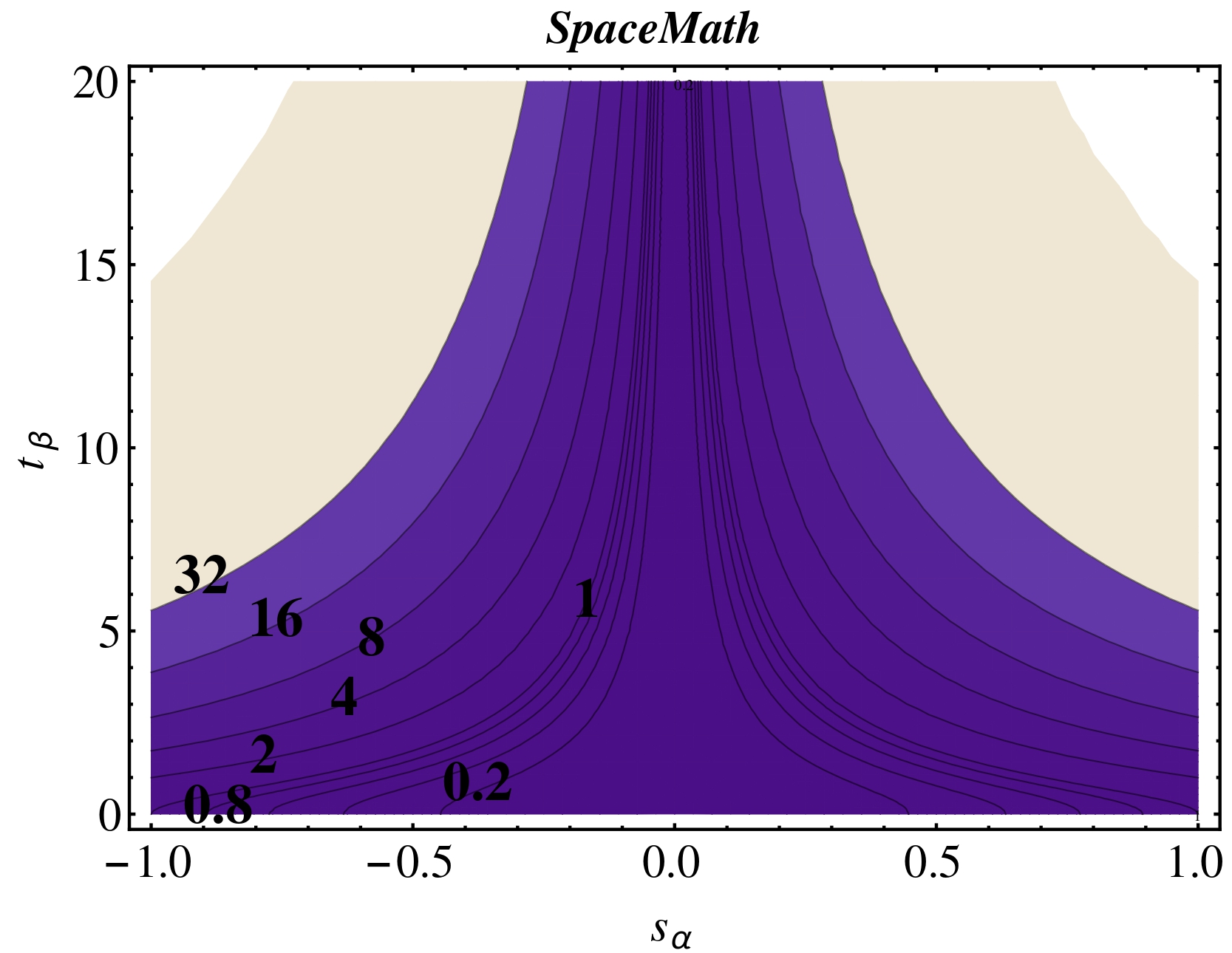}
		\caption{Contours of $\Gamma(h \to b \bar b) / \Gamma(h_{SM} \to b \bar b)$ 
			for the SM-like Higgs boson as a function of $\sin \alpha$ and 
			$\tan \beta$ in Type 2 THDM. Left: figure taken from \cite{Craig:2012vn} and Right: figure generated by \texttt{SpaceMath v.2.0}. 
		}
		\label{fig:widthbb2}
	\end{figure}

\begin{figure}[!h] 
		\centering
		\includegraphics[width=5cm]{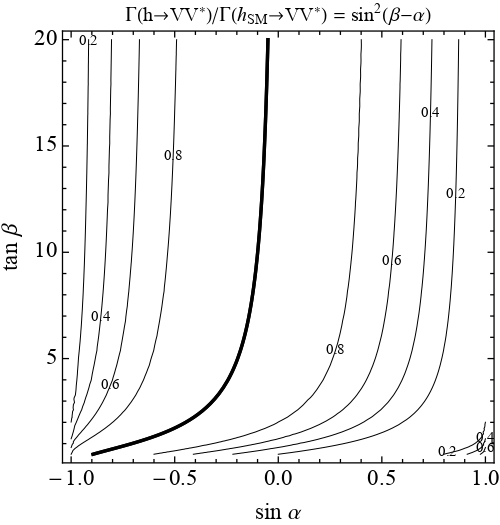} 
		\includegraphics[width=6.5cm]{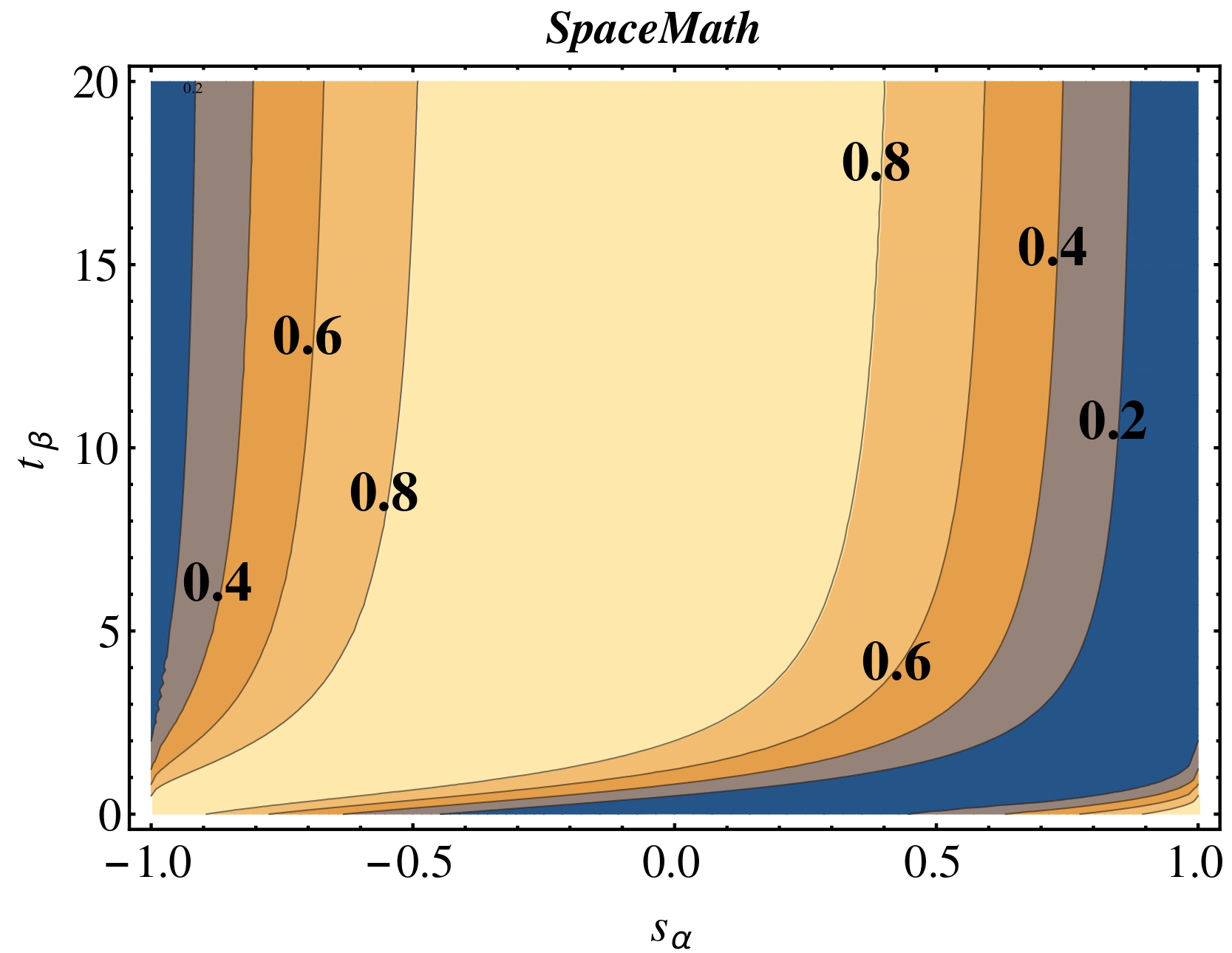}
		\caption{Contours of 
			$\Gamma(h \to VV^{\star}) / \Gamma(h_{SM} VV^{\star})$ 
			for the SM-like Higgs boson as a function of $\sin \alpha$ and 
			$\tan \beta$ in any of the THDMs. Left: figure taken from \cite{Craig:2012vn} and Right: figure generated by \texttt{SpaceMath v.2.0}. 
		}
		\label{fig:widthVV}
	\end{figure}

	\noindent We have defined $\texttt{Sa\ensuremath{\equiv}sin\ensuremath{\alpha}, Tb\ensuremath{\equiv}\ensuremath{\tan\beta}, Cb\ensuremath{\equiv}\ensuremath{\cos\beta}, Sb\ensuremath{\equiv\sin\beta}}$
	are free parameters of THDM-I, -II. In addition, we have used the relations $\tan\beta=\frac{\sin\beta}{\cos\beta}$,
	$\sin(\beta-\alpha)=\sin\beta\cos\alpha-\cos\beta\sin\alpha$. The commands $\texttt{kb}$ and $\texttt{kV}$ can be directly evaluated by introducing values for $\texttt{Sa, Tb, Cb}$, or since
	$\texttt{SpaceMath}$ is hosted in \texttt{Mathematica}, we can use its commands
	to graph. For this example we use:
	\begin{itemize}
		\item $\texttt{ContourPlot[kb[\texttt{ghbb[Sa,Tb,Cos[ArcTan[Tb]]}]]\textasciicircum2,\{Sa,-1,1\},\{Tb,0,20\}]},$
		\item 
		$\texttt{ContourPlot[kV[ghVV[Tb,Cos[ArcTan[Tb]],Sin[ArcTan[Tb]],Sa]]\textasciicircum2,\{Sa,-1,1\},\{Tb,0,20\}]},$
	\end{itemize}
	which generate the graphs displayed in Figs. \ref{fig:widthbb1}, \ref{fig:widthbb2} and \ref{fig:widthVV}. The codes that generate these graphs can be found in the $\texttt{"Examples"}$ directory, whose path is:\\ \texttt{\$SpaceMath/Examples/Validation\_RX/SPACEMATH\_RX-Validation-THDM.nb}\\
	 or click on the link \texttt{"Examples"} once \texttt{SpaceMath} was loaded.

\begin{figure}[!h]
		\centering
		\subfigure[]{{\includegraphics[scale=0.3]{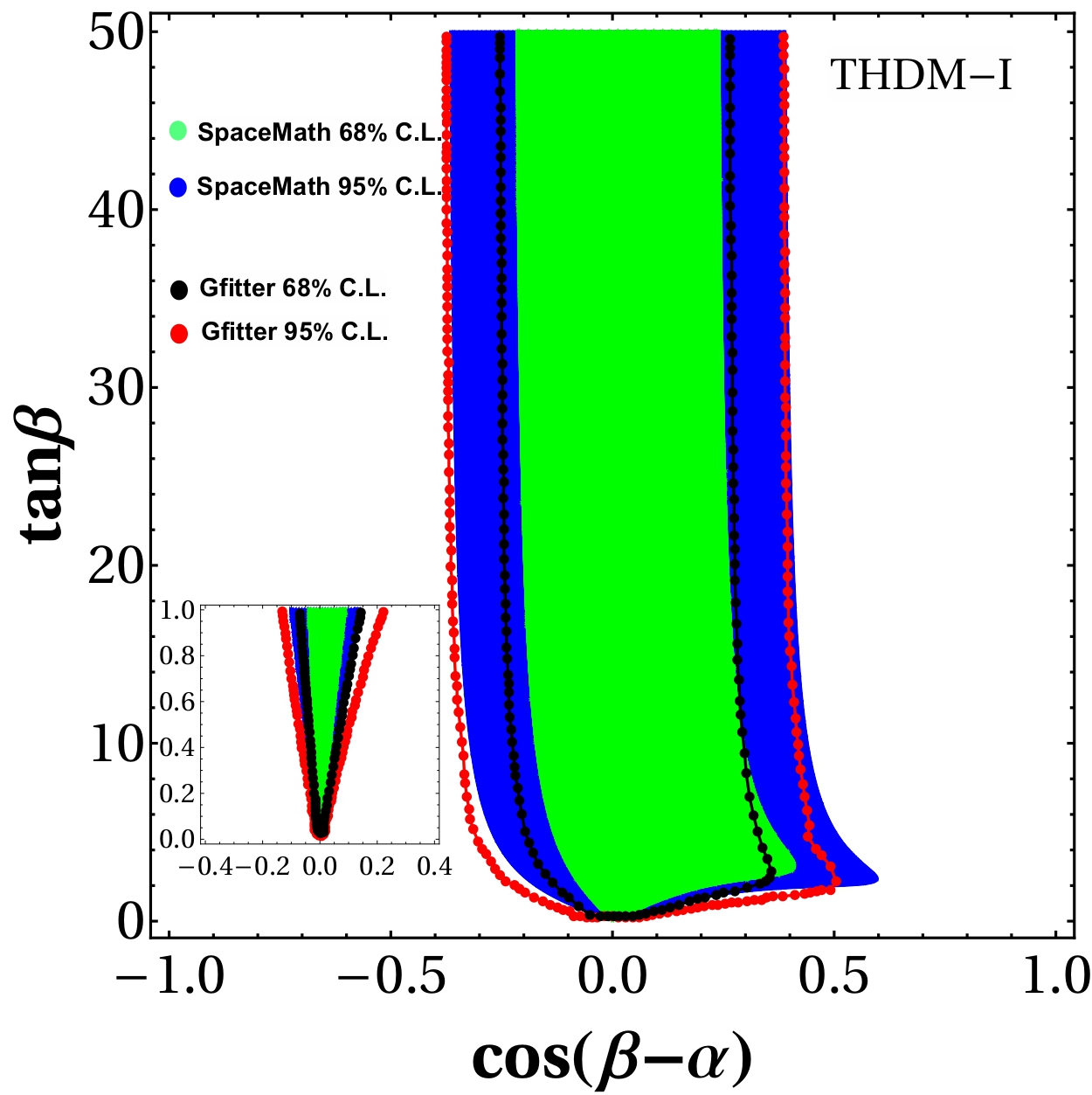}}\label{a}}
		\subfigure[]{{\includegraphics[scale=0.3]{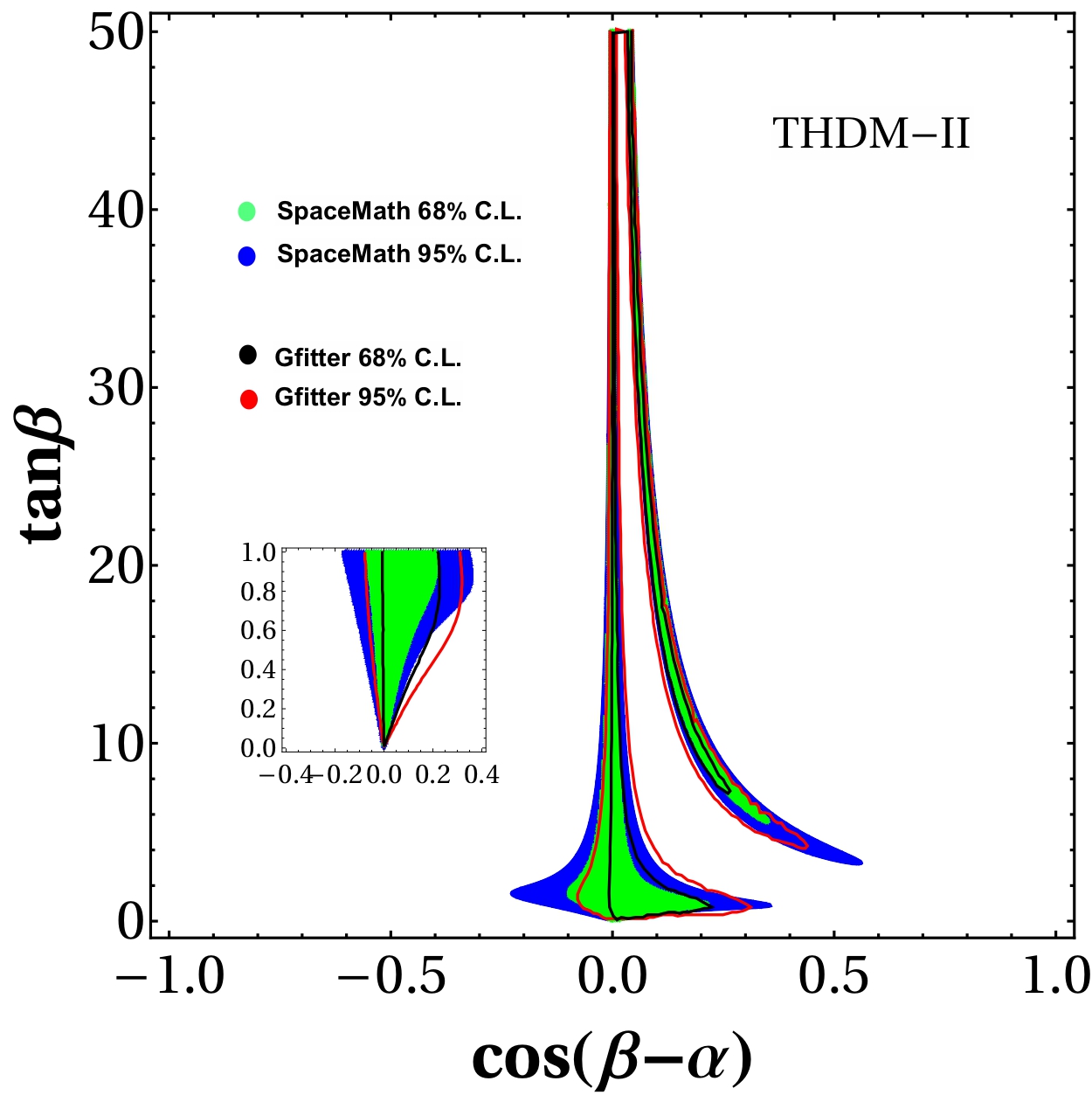}}\label{b}}
		\subfigure[]{{\includegraphics[scale=0.3]{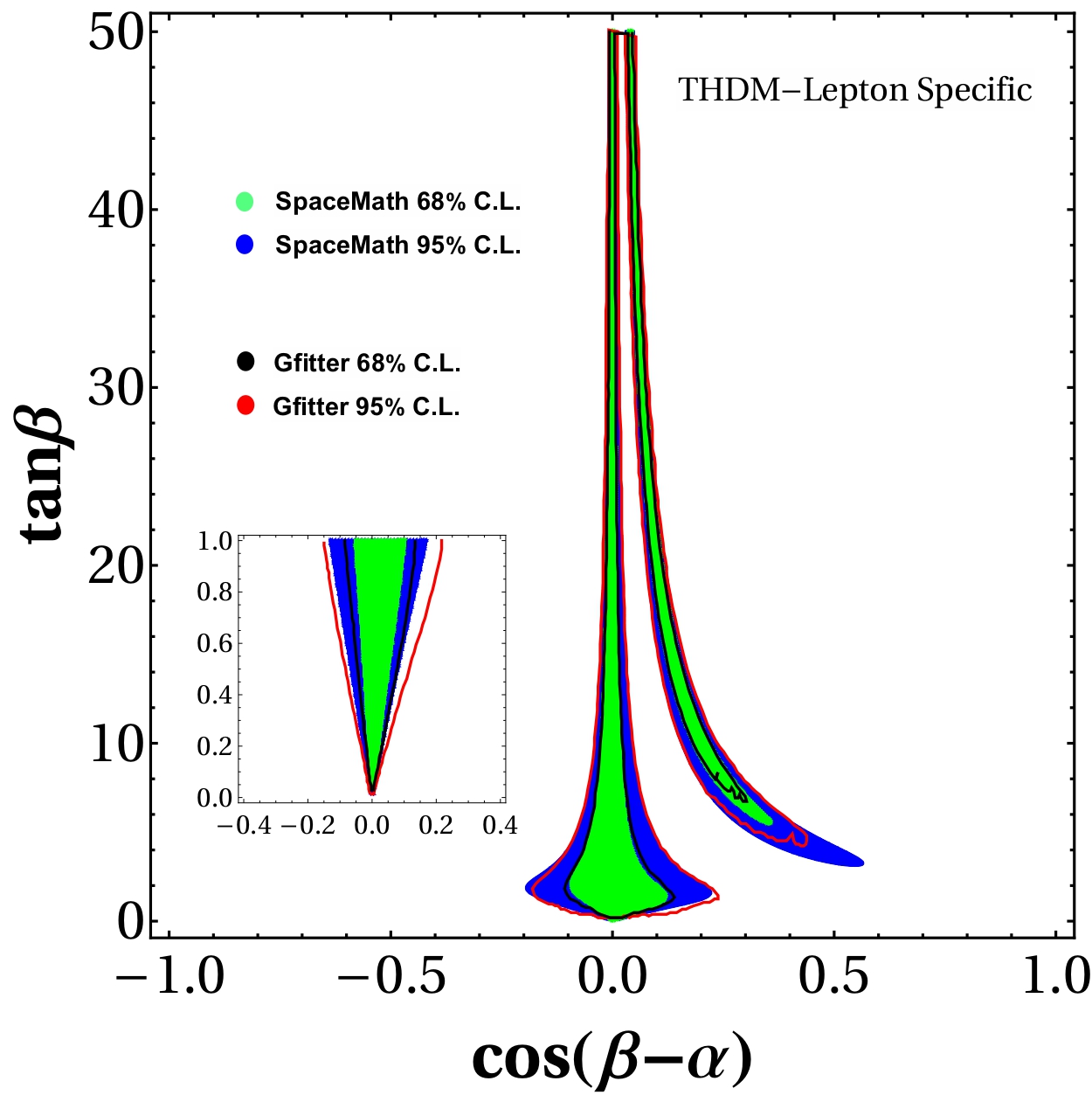}}\label{c}}
		\subfigure[]{{\includegraphics[scale=0.3]{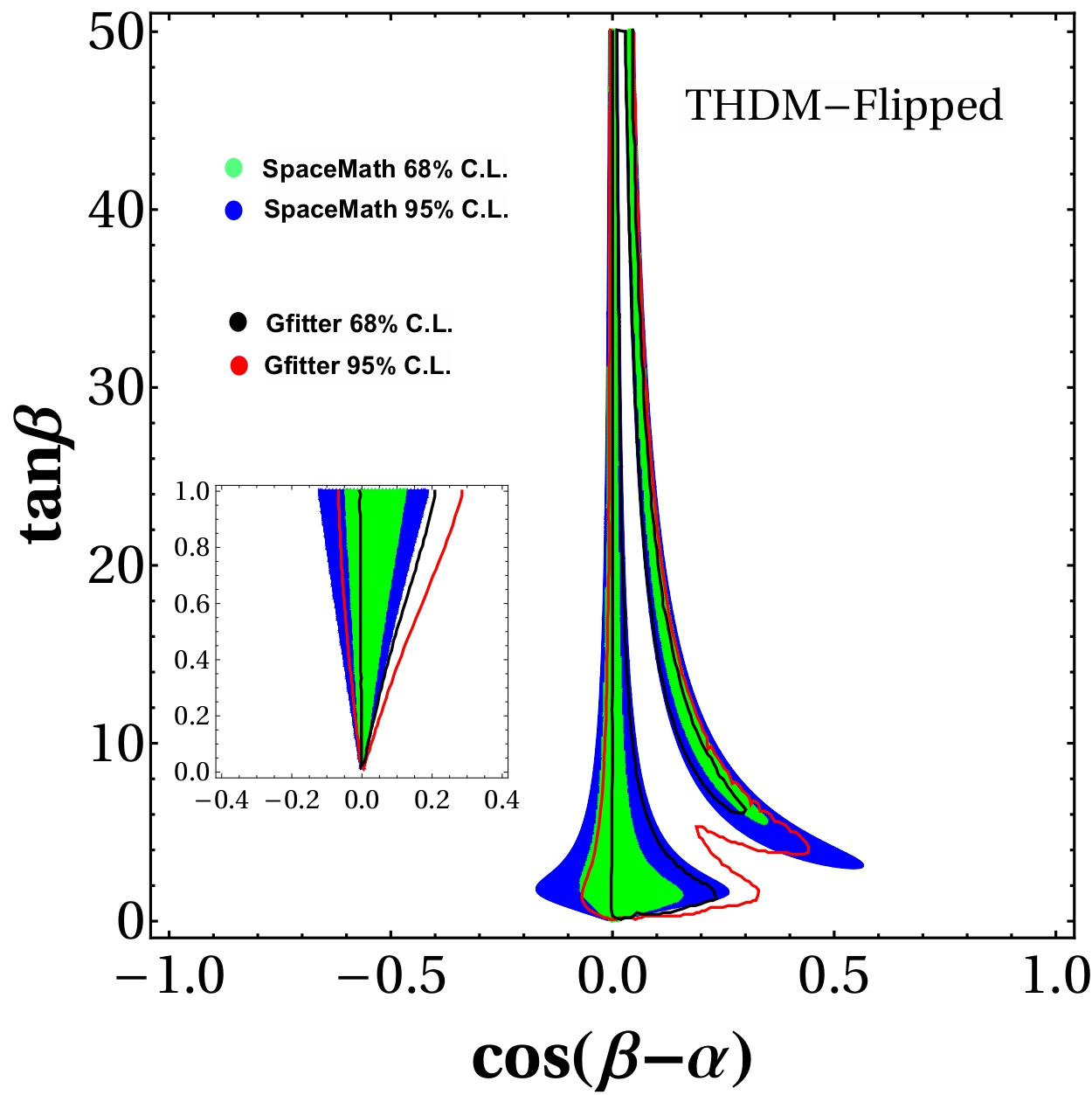}}\label{d}}
		\caption{Plane $\cos(\beta-\alpha)-\tan\beta$ for different versions of THDM's: (a) Type I, (b) Type II, (c) Lepton Specific, (d) Flipped. The plots were generated  in \texttt{SpaceMath v.2.0}.\label{tb_cab}}
	\end{figure}

	In addition, we also show in Fig. \ref{tb_cab} the THDM-I, -II, Lepton Specific and Flipped parameter spaces in the $\cos(\beta-\alpha)-\tan\beta$ plane. Again, couplings are shown in Table \ref{AcoplosTHDM}. We compare our results with the ones reported by authors of Ref. \cite{Haller:2018nnx}. In these graphs we perform a $\chi^2$ test which is defined as follows:
	\begin{equation}
		\chi^2=\sum_{i=1}^n\left(\frac{O_i-E_i}{\sigma_i}\right)^2,
	\end{equation}
	where $O_i$ and $E_i$ are the observed and expected values, respectively, and $\sigma_i$ indicates uncertainty. The command for plot these figures is: 
	
	\begin{center}\resizebox{\textwidth}{!}{%
			\fbox{\begin{minipage}{\dimexpr\textwidth+55\fboxsep-1.5\fboxrule\relax}	
					\centering 
					$ \texttt{Chi2Rx95[ghtt[-ArcCos[Cab] + ArcTan[tb], tb],ghbb[-ArcCos[Cab] + ArcTan[tb], tb], }$\\
					$\texttt{ghtautau[-ArcCos[Cab] + ArcTan[tb], tb], ghZZ[Sqrt[1 - Cab\textasciicircum2]],ghWW[Sqrt[1 - Cab\textasciicircum2]], 0, 2000, Cab, tb]}$;
					\\ 
					$ \texttt{Chi2Rx68[ghtt[-ArcCos[Cab] + ArcTan[tb], tb],ghbb[-ArcCos[Cab] + ArcTan[tb], tb], }$\\
					$\texttt{ghtautau[-ArcCos[Cab] + ArcTan[tb], tb],ghZZ[Sqrt[1 - Cab\textasciicircum2]],ghWW[Sqrt[1 - Cab\textasciicircum2]], 0, 2000, Cab, tb]}$
		\end{minipage}}}
	\end{center}
	
	Complete instructions can be found at: \\
	\texttt{\$SpaceMath/Examples/Validation\_RX/SPACEMATH\_RX-Validation-THDM-Chi2Rx.nb}.    
	
	\begin{center}
		\begin{table}[H]
			\caption{THMD's $hff$ and $hVV$ couplings.\label{AcoplosTHDM}}
			\centering
			\resizebox{\textwidth}{!}{%
				\begin{tabular}{ccccc}
					\hline 
					Coupling& THDM-I & THDM-II & THDM-Lepton Specific & THDM-Flipped\tabularnewline
					\hline 
					\hline 
					$hVV$ & $\sin(\beta-\alpha)$ & $\sin(\beta-\alpha)$ & $\sin(\beta-\alpha)$ & $\sin(\beta-\alpha)$\tabularnewline
					\hline 
					$hu_{i}u_{i}$ & $\cos\alpha/\sin\beta$ & $\cos\alpha/\sin\beta$ & $\cos\alpha/\sin\beta$ & $\cos\alpha/\sin\beta$\tabularnewline
					\hline 
					$hd_{i}d_{i}$ & $\cos\alpha/\sin\beta$ & $-\sin\alpha/\cos\beta$ & $\cos\alpha/\sin\beta$ & $-\sin\alpha/\cos\beta$\tabularnewline
					\hline 
					$h\ell_{i}\ell_{i}$ & $\cos\alpha/\sin\beta$ & $-\sin\alpha/\cos\beta$ & $-\sin\alpha/\cos\beta$ & $\cos\alpha/\sin\beta$\tabularnewline
					\hline 
				\end{tabular}
				\par
			}
		\end{table}
	\end{center}				
	
	We can observe slight differences between the graphs generated via \texttt{SpaceMath v.2.0} and those of the \texttt{Gfitter} group, this is due to two sources: 1) The experimental data that \texttt{SpaceMath} considers are the most recent and 2) the \texttt{Gfitter} team includes all production modes of the Higgs boson. Here, it is worth mentioning that even though \texttt{SpaceMath v.2.0}  only has gluon fusion production implemented, our results are highly similar, this may be because it is the dominant channel for the production of the higgs boson.
	
	Finally, we shown in Table \ref{hdecayTHDMtype1} a comparison between our numerical evaluations and those made via \texttt{HDecay} package \cite{Djouadi:2018xqq}, which the branching ratios of the Higgs boson decaying to pair of particles ($b\bar{b}$, $s\bar{s}$, $c\bar{c}$, $t\bar{t}$, $\tau^+\tau^-$, $\mu^+\mu^-$, $gg$, $\gamma\gamma$, $Z\gamma$, $W^+W^-$, $ZZ$) in the theoretical framework of the THDM-I are shown. Again, the Feynman rules needed for evaluations are shown in Table \ref{AcoplosTHDM}, where it can be seen that only two parameters are introduced. We take the same inputs for these free THDM-I parameters as in Ref. \cite{Djouadi:2018xqq}, namely,
	\begin{itemize}
		\item $\tan\beta$= 1.29775,
		\item $\alpha$=-0.684653,
	\end{itemize}	  
	and we also consider a Higgs boson mass of $m_{h}$=125.09 GeV.
	
	\begin{center}
		\begin{table}[H]
			\caption{Comparison of numerical evaluations computed by \texttt{SpaceMath v1.0} and \texttt{HDecay}. The theoretical framework used is the THDM-I, whose Feynman rules are shown in Table \ref{AcoplosTHDM}. Results in brackets are those generated via \texttt{SpaceMath V.2.0}. \label{hdecayTHDMtype1}}
			\centering
			\resizebox{\textwidth}{!}{%
				\begin{tabular}{llllll}
					\hline \rule{0pt}{3ex} 
					$\mathcal{BR}(h\rightarrow b\bar{b})$ & $\mathcal{BR}(h\rightarrow\tau\tau)$ & $\mathcal{BR}(h\rightarrow\mu\mu)$ & $\mathcal{BR}(h\rightarrow s\bar{s})$ & $\mathcal{BR}(h\rightarrow c\bar{c})$ & $\mathcal{BR}(h\rightarrow t\bar{t})$ \tabularnewline
					0.6080 (\textit{0.6080}) & 0.6542 (\textit{0.6542})$\times10^{-1}$ & 0.2316 (\textit{0.2316})$\times10^{-3}$ & 0.2294 (\textit{0.2294})$\times10^{-3}$ & 0.2653(\textit{0.2653})$\times10^{-1}$  & 0 (\textit{0}) \tabularnewline				 
					\hline \rule{0pt}{3ex} 
					$\mathcal{BR}(h\rightarrow gg)$ & $\mathcal{BR}(h\rightarrow\gamma\gamma)$ & $\mathcal{BR}(h\rightarrow Z\gamma)$ & $\mathcal{BR}(h\rightarrow WW)$ & $\mathcal{BR}(h\rightarrow ZZ)$ & \tabularnewline
					0.7041 (\textit{0.7041})$\times10^{-1}$ & 0.2126 (\textit{0.2126})$\times10^{-2}$ & 0.1458 (\textit{0.1458})$\times10^{-2}$ & 0.2005 (\textit{0.2005}) & 0.2507 (\textit{0.2507})$\times10^{-1}$  & \tabularnewline				 
					\hline \rule{0pt}{3ex} 
					$\mathcal{BR}(h\rightarrow AA)$ & $\mathcal{BR}(h\rightarrow AZ)$ & $\mathcal{BR}(h\rightarrow W\pm h\mp)$ & $\mathcal{BR}(h\rightarrow h+h-)$ & $\Gamma_{h}^{\textrm{tot}}$ & \tabularnewline			    	 
					0 (\textit{0}) & 0 (\textit{0})& 0 (\textit{0})& 0 (\textit{0})&  0.4248 (\textit{0.4248})$\times10^{-2}$ GeV  & \tabularnewline				 
					\hline															 
				\end{tabular}
				\par
			}
		\end{table}     
	\end{center}
	In Table \ref{hdecayTHDMtype1}, the quantities in brackets are the results generated via \texttt{SpaceMath}. We observe that our results are identical to those \texttt{HDecay}, which is to be expected since we actually reproduced the relevant expressions of the decay widths of the Higgs boson reported in Ref. \cite{Djouadi:2005gi}.

	\section{Conclusions}\label{SeccionV}
	We have introduced a \texttt{Mathematica} package called \texttt{SpaceMath v.2.0} which generates parameter spaces of Standard Model extensions that are in agreement with current experimental measurements. The physical observables considered in this version are LHC Higgs boson data (and its projections for HL-LHC and HE-LHC) and Lepton Flavor Violating processes. \texttt{SpaceMath v.2.0} complements the previous version by implementing Machine Learning algorithms, namely, \textit{Linear Regression, Decision Trees, Gradiant Boosted Trees, Neural Networks and Gaussian Process}, which will help us predict Benchmark Points to be used directly in evaluations of calculataions of physical observables. We show in detail how \texttt{SpaceMath v.2.0} works by appliying it to the Two-Higgs Doublet Model of type III. 
	\section*{Acknowledgments}
	The work of Marco A. Arroyo-Ure\~na and T. Valencia-P\'erez is supported by ``Estancias posdoctorales por M\'exico (CONAHCYT)" and ``Sistema Nacional de Investigadores" (SNI-CONAHCYT). We also thank Dr. Olga Felix and her team for the computer resources and technical advice. T.V.P thanks to Dr. Myriam Mondragón for her valuable suggestions during the development of this research work.
	\appendix
	
	\section{Remote connection}\label{cluster}
	Requirements to remote connection:
	\begin{itemize}
		\item \texttt{Mathematica} version: 12.0.++
		\item PowerShell (windows).
	\end{itemize}
	Steps to connect to server $``Negrito"$.
	\begin{enumerate}
		\item Open a terminal and type \texttt{\$ ssh spacemathuser@148.228.14.13 -Y}.
		\item Enter password: \texttt{spacemath}
		\item Type \texttt{mathematicaX}, where \texttt{X=12,13} represents the \texttt{Mathematica} version. 
		\item Enjoy \texttt{SpaceMath v.2.0} package.
	\end{enumerate}


	
	\newpage

\end{document}